\documentclass[12pt]{iopart}
\usepackage{iopams} 
\usepackage{epsfig} 
\begin{document}

\title[Confinement-induced resonances in arbitrary quasi-1D traps]
{Confinement-induced resonances for a two-component ultracold
atom gas in arbitrary quasi-one-dimensional traps}
 
\author{ V Peano, M Thorwart, C Mora and R Egger}

\address{Institut f\"ur Theoretische Physik, 
Heinrich-Heine-Universit\"at D\"usseldorf, D-40225 D\"usseldorf, Germany}

\submitto{\NJP}

\begin{abstract}
We solve the two-particle s-wave
scattering problem for ultracold atom gases confined in arbitrary
quasi-one-dimensional trapping potentials,
allowing for two different atom species. As a consequence, 
the center-of-mass and relative degrees of freedom do not
factorize. We derive bound-state solutions 
and obtain the general
scattering solution, which exhibits several resonances in the 1D scattering
length induced by the confinement. 
  We apply our formalism to two experimentally relevant cases:  (i) 
 interspecies scattering in a two-species mixture, 
 and (ii) the two-body problem for  a single species in a
 non-parabolic trap. 
\end{abstract}
\pacs{03.65.Nk, 03.75.Mn, 34.50.-s}
\maketitle
\section{Introduction}
A strongly interacting ultracold atom gas displays interesting
features of a correlated quantum  many-body system 
when its dynamics is confined to one dimension \cite{Petrov04}. 
The presence of a transverse confining potential has been 
shown to induce characteristic resonances in the coupling constant of 
the two-particle s-wave scattering process 
\cite{olshanii,bergeman,Moore}, which have
become known as confinement-induced resonances (CIR).  
The existence of the CIR has been revealed under the simplifying
assumption of a transverse parabolic confinement potential with 
length scale $a_\perp$ and for the case that the two
scattering atoms belong to the same species 
\cite{olshanii,bergeman,Moore}. 
In this case, the center-of-mass (COM) 
  and relative coordinates of the two particles can be  
separated, allowing to factorize the problem into single-particle
problems. At low temperatures, only the COM ground state is occupied,
the decoupled COM motion can be disregarded,
and the two-body problem can be solved exactly within the pseudopotential
approximation. The result is that there is  exactly one 
 bound state for any 3D scattering length $a$. In the limit of small 
 binding energy, the particles are 
 tightly bound in the lowest-energy transverse state
 and form a very elongated dimer. The appearance of such a bound state 
 is purely due to the confinement, since for $a<0$ no dimer
 is formed in free space.  In the opposite limit of large binding
 energies,  the dimer becomes spherically symmetric. 
 In this regime, the  confinement is not effective, 
 and the free-space result is recovered. Moreover, 
 a unitary equivalence exists between the Hamiltonian and 
 its projection onto those channels which are perpendicular 
 to the one with lowest energy. As a consequence, 
 to each bound state corresponds a bound state of the closed channels,
 which then causes the CIR \cite{bergeman}. It occurs at a universal value 
  of the ratio $a_\perp/a={\cal C}=-\zeta(1/2,1) \simeq 1.4603$, where
  $\zeta$ is the Hurvitz zeta function. 
 The influence of the CIR has also been studied for the three-body~\cite{Mora05}
 and the four-body problem~\cite{Mora06} in the presence of confinement. 
 In particular, the solution of the four-body problem  completely determines
the corresponding quasi-1D many-body BCS-BEC crossover 
phenomenon \cite{Mora06}.
Recently, the existence of the confinement-induced molecular bound
 state in a  quasi-1D Fermionic $^{40}$K atom gas confined in an optical trap 
  has been reported~\cite{Moritz}. 
 By using rf spectroscopy, the binding energy of the dimer has been
 measured as a function of the scattering length, with quantitative
 agreement to the results of Ref.~\cite{bergeman}.  However, the existence 
 of the CIR in the scattering states remains to be observed.   
 
 Although the analytical results for the parabolic confinement are
 instructive, realistic traps 
 for matter waves frequently have non-linear potential forms, see for
 instance Ref.\  \cite{Peano} for a particular example of a trap on
 the nanometer scale.  
 To give another example, for the problem of tunneling of a macroscopic number of
 ultracold atoms between two stable states of a trapping potential,
 the nonlinearity clearly is crucial.  
 Hence  generalization to the non-parabolic case is
 desirable and provided in this work. In addition, we consider
 traps with two different
 species of atoms. Note that sympathetic cooling techniques require to
 study this case. Different trap frequencies may arise for different
 atom species, e.g., because of different atom
masses or  different magnetic quantum numbers. Here we  obtain 
general expressions for the bound-state energies and scattering
resonances when the COM
and the relative degrees of freedom do not decouple anymore. 
In the parabolic limit and for
 intraspecies scattering, we recover well-known results
\cite{olshanii,bergeman}. For the general 
case, we show that more
than one CIR may appear, and that it depends on the symmetry properties of the
confining potential how many resonances occur. We
apply our formalism to two experimentally relevant cases:  (i)
 interspecies scattering in a two-species mixture of quantum degenerate Bose
 and Fermi gases in an optical trap, and (ii)  a single species cloud in a
 magnetic trap, taking into account non-parabolic corrections due
 to a longitudinal magnetic field suppressing Majorana spin
 flips.
  
As we will discuss below in more detail, the CIR has a close
similarity to the well-known Feshbach resonance \cite{timmer}, which
arises if the Hilbert space can be
divided into open and closed channels coupled together by a short-range
 interaction. Due to this small but finite
 coupling, two incoming particles initially in the open
channel  visit the closed channels  during the
 scattering process. 
If a bound state with  energy close to the continuum threshold 
exists, such a process is highly enhanced  and a resonance results.

The paper is organized as follows: In Sec.\  \ref{model}, we present
the general formalism.  Section  \ref{sec.boundstates} presents the
bound-state solution, while Sec.\ \ref{sec.scatter} contains the analysis for
the scattering solutions, including the analogy to Feshbach
resonances. In Sec.\  \ref{sec.parabolic} we discuss the special case
of harmonic confinement, and in Sec.\ \ref{sec.nonlinear} a particular
example of a non-parabolic confinement is illustrated. Finally, we 
conclude in Sec.\ \ref{sec.conclusio}.  Technical
details have been delegated to two Appendices.
We set $\hbar=1$ throughout this paper.

\section{The two-body problem \label{model}}

Let us consider the general case of two different atomic species with mass
 $m_1$ and $m_2$. We denote the particle coordinates by  
${\bf x}_{i}=({\bf x}_{\perp,i},z_{i})$ and their momenta by  
${\bf p}_{i}=({\bf p}_{\perp,i},p_{\parallel,i})$. 
Different atoms may experience a different transversal confinement 
potential $V_{i}({\bf x}_{\perp,i})$. 
For ultracold atoms, only low-energy s-wave scattering is relevant,
and the interaction between unlike atoms (and similarly,
also the interaction between the same atoms)   can be described by 
a Fermi-Huang pseudopotential $V(|{\bf x}_1-{\bf x}_2|)$. 
Then the relevant  Hamiltonian for two different atoms is given by 
\begin{equation}
H=\frac{{\bf p}_1^2}{2m_1}+\frac{{\bf p}_2^2}{2m_2}+
V_1({\bf x}_{\perp,1})+
V_2({\bf x}_{\perp,2})+V(|{\bf x}_1-{\bf x}_2|)\, .
\end{equation} 
The pseudopotential has the standard form
\begin{equation}\label{pseudo}
V({\bf r}) = \frac{2\pi a}{\mu} \delta({\bf r}) 
\frac{\partial}{\partial r} r \, ,
\end{equation}
where $\mu=m_1 m_2/(m_1+m_2)$ is the reduced mass and $a$ the  
3D scattering length. This allows to characterize the 
two-body interaction by the parameter $a$ only. 
The validity of the pseudopotential approach 
has been verified numerically for finite-range potentials 
in Ref.~\cite{bergeman}.
For further convenience, we transform to the relative/COM coordinates and
momenta given by 
 ${\bf r}=({\bf r}_\perp,z)$, ${\bf R}=({\bf R}_\perp,Z)$
and ${\bf p}=({\bf p}_\perp,p_\parallel)$, 
${\bf P}=({\bf P}_\perp,P_\parallel)$, respectively.
This can be done by the canonical transformation
\begin{equation}
\left(
\begin{array}{c}
{\bf R}\\{\bf r}\\{\bf P}\\{\bf p}
\end{array}
\right)
=\frac{1}{M}
\left(
\begin{array}{cccc}
m_1&m_2&0&0\\M&-M&0&0\\0&0&M&M\\0&0&m_2&-m_1
\end{array}
\right)
\left(
\begin{array}{c}
{\bf x}_1\\{\bf x}_2\\{\bf p}_1\\{\bf p}_2
\end{array}
\right) \, , 
\end{equation}
where $M=m_1+m_2$. Since the confinement is assumed to be purely transversal,
 the longitudinal COM coordinate
 $Z$ is free and decouples from the other degrees of
freedom. Hence we eliminate it by transforming into the longitudinal 
COM rest frame, where
the state $|\Psi\rangle$ of the system  is
 determined
 by the set of coordinates $({\bf x}_{\perp,1},{\bf x}_{\perp,2},z)$ 
or, alternatively, by $({\bf R}_{\perp},{\bf r})=({\bf R}_{\perp},{\bf
r}_\perp,z)$. The transformed Hamiltonian takes the form
\begin{equation}
H=H_\parallel+H_{\perp,1}+H_{\perp,2}+V\, , 
\end{equation}
where
\begin{equation}
H_\parallel=\frac{p_\parallel^2}{2\mu},\quad H_{\perp,i}=
\frac{{\bf p}_{\perp,i}^2}{2m_{i}}
+V_{i}({\bf x}_{\perp,i}) \, .
\end{equation}
For a more compact notation, we introduce the non-interacting Hamiltonian 
$H_0 =H-V$ and denote its eigenstates by  
\begin{equation}
|k,\lambda_1,\lambda_2\rangle=e^{-ikz}\psi^{(1)}_{\lambda_1}({\bf x}_{\perp,1})
\psi^{(2)}_{\lambda_2}({\bf x}_{\perp,2})\, ,
\end{equation}
where $\psi^{(i)}_{\lambda_i}$ are single-particle 
eigenstates of $H_{\perp,i}$ for
eigenvalue $E^{(i)}_{\lambda_i}$. Correspondingly, the 
two-particle Schr\"odinger equation is given by 
\begin{equation}
\left(H_0- E\right)\Psi({\bf R}_\perp,{\bf r})
 = -V({\bf r})\Psi({\bf R}_\perp,{\bf r}) \, .
\end{equation}
The pseudopotential (\ref{pseudo}) can be enforced by the
 Bethe-Peierls boundary condition
\begin{equation}
\Psi({\bf R}_\perp,{\bf r}\to 0) \simeq \frac{f({\bf R}_\perp)}{4\pi
r}\left(1-\frac{r}{a}\right)
\label{bc} \, ,
\end{equation}
leading to the  inhomogeneous   Schr\"odinger equation
\begin{equation}
\left(H_0- E\right)\Psi({\bf R}_\perp,{\bf r})
 = 
\frac{f({\bf R}_\perp)}{2\mu}\delta({\bf r}).
\end{equation} 
The solution of this equation can be formally obtained 
 in terms of a solution of the homogeneous Schr\"odinger 
 equation, $(H_0- E)\Psi_0=0$, 
and the  Green's function $G_E=(H_0-E)^{-1}$, 
\begin{equation}
\Psi({\bf R}_\perp,{\bf r})=\Psi_0({\bf R}_\perp,{\bf r})+
\int d{\bf R}^{\prime}_\perp\,
G_E({\bf R}_\perp,{\bf r};{\bf R}^{\prime}_\perp,0)
\frac{f({\bf R}^{\prime}_\perp)}{2\mu}\, .\label{Gensol}
\end{equation}
To determine $f({\bf R}_\perp)$, we substitute 
 Eq.\  (\ref{Gensol})  into
Eq. (\ref{bc}) and find the integral equation
\begin{equation}
-\frac{f({\bf R}_\perp)}{4\pi a}=\Psi_0({\bf R}_\perp,0)+
\int d{\bf R}_\perp^\prime
 \zeta_{E}({\bf R}_\perp,{\bf R}_\perp^\prime)
f({\bf R}_\perp^\prime),
\label{inteq}
\end{equation}
where we have defined the regularized integral kernel
\begin{equation}
\zeta_{E}({\bf R}_\perp,{\bf R}_\perp^\prime)=\lim_{r\rightarrow0}
\frac{1}{2\mu}\Big(G_E({\bf R}_\perp,{\bf r}; {\bf R}'_\perp, 0)
-\delta({\bf R}_\perp-{\bf R}_\perp^\prime) \frac{\mu}{2\pi r}\Big)
.\label{kernel}
\end{equation}
In Eqs.\  (\ref{Gensol}) and (\ref{inteq}), 
$\Psi_0$ can be expressed 
as a superposition of single-particle eigenstates 
$|k,\lambda_1,\lambda_2\rangle$ with
 \begin{equation}
 \frac{k^2}{2\mu}+E^{(1)}_{\lambda_1}+E^{(2)}_{\lambda_2}=E\,.
\label{fE}
\end{equation} 
We refer to the set of states with the same transverse occupation 
numbers $\lambda_i$ but arbitrary longitudinal relative momentum as a {\it
scattering channel\/} or, simply, {\it channel\/}. 
 Each channel has a minimum energy given by 
$E^{(1)}_{\lambda_1}+E^{(2)}_{\lambda_2}$.
 Since the interaction is short-ranged, only 
states fulfilling Eq.\ (\ref{fE}) appear in the asymptotic solution. 
For each {\sl open channel}, $E>E^{(1)}_{\lambda_1}+E^{(2)}_{\lambda_2}$,
such that there are (at least) two such states  having opposite momenta. 
For $E$ just above $E^{(1)}_0+E^{(2)}_0$, there exists one open
 channel only. The corresponding solution given by Eq.\ (\ref{Gensol})
 describes the 
scattering of two particles initially occupying the transverse ground-state.
During the scattering process, the particles  populate
closed channels, but afterwards
 return into the single available
 open channel (quasi-1D picture).
For   $E<E^{(1)}_0+E^{(2)}_0$, all channels are closed and
only bound-state solutions are possible. These are given by  
 Eq.\ (\ref{Gensol}) with $\Psi_0({\bf R}_\perp,{\bf r})=0$.  
In the following, we consider both classes of solutions in more detail. 

\section{Bound-state solutions\label{sec.boundstates}}
Let us consider the situation when all channels are closed
and only bound states may occur.
 We define the binding energy of the bound states
 as
\begin{equation}
E_B = E_0-E>0 \, , \label{E_B}
\end{equation}
where $E_0 = E^{(1)}_0+E^{(2)}_0$ is the ground-state energy of
$H_0$. To find bound states, we  
diagonalize the operator 
$\zeta_{E}({\bf R}_\perp, {\bf R}^\prime_\perp )$
defined in Eq.~(\ref{kernel}), where Eq.~(\ref{inteq}) 
yields the condition
\begin{equation}
-\frac{f({\bf R}_\perp)}{4\pi a}=
\int d{\bf R}_\perp^\prime
 \zeta_{E}({\bf R}_\perp,{\bf R}_\perp^\prime)
f({\bf R}_\perp^\prime)\, .
\label{eigeq}
\end{equation}
For given $a$, bound states with binding energy 
$E_B=E_0-E$    follow as solution of this eigenvalue problem.
The bound-state wave function follows by inserting the corresponding 
eigenvector $f({\bf R}_\perp)$ into Eq.\ (\ref{Gensol}) with
  $\Psi_0({\bf R}_\perp,{\bf r})=0$. 
In order to find a representation of 
 $\zeta_{E}({\bf R}_\perp, {\bf R}^\prime_\perp )$ allowing for
straightforward analytical or numerical diagonalization, we  use
\begin{equation}
 G_E({\bf R}_\perp,{\bf r};{\bf R}^\prime_\perp,0)=
\int_0^\infty dt\,e^{Et}
G_t({\bf R}_\perp,{\bf r};{\bf R}_\perp^\prime,0)
\label{GE}\,,
\end{equation}
with the imaginary-time evolution operator 
\begin{equation}
G_t({\bf R}_\perp,{\bf r};{\bf R}^\prime_\perp,0)=
\langle{\bf R}_\perp,{\bf r}|\exp[- H_0t]|{\bf
R}^\prime_\perp,0\rangle 
\end{equation}
 for  $H_0$.  The time evolution operator  $\exp[- H_0t]$ can be factorized 
into the product 
$\exp[-H_\parallel t]\exp[-H_{1,\perp} t]\exp[-H_{2,\perp} t]$.
The corresponding factors in $G_t$ are
\begin{equation}
\langle z|\exp[-H_\parallel t]|z'\rangle=
\left(\frac{\mu}{2\pi t}\right)^{1/2}e^{-(z-z')^2\mu/2t} \, 
\end{equation}
for the relative longitudinal coordinates  and 
\begin{equation}
\langle{\bf x}_{i,\perp}|\exp[-H_{i,\perp} t]|{\bf x}_{i,\perp}'\rangle
=\sum_{\lambda} e^{-E^{(i)}_\lambda t}\psi^{(i)}_\lambda({\bf x}_{i,\perp})
\bar{\psi}^{(i)}_\lambda({\bf x}_{i,\perp}^\prime)\, 
\end{equation}
for the transverse coordinates
(the bar denotes  complex conjugation). 
Thus $G_t$ can be expressed in terms of the set of coordinates 
$({\bf x}_{\perp,1},{\bf x}_{\perp,2},z)$ as
\begin{equation}
G_t({\bf R}_\perp,{\bf r};{\bf R}^\prime_\perp,0)=
\sqrt{\frac{\mu}{2\pi t}}e^{-z^2\mu/2t}\prod_{i=1,2}\sum_\lambda
e^{-E^{(i)}_{\lambda}t}
\psi_{\lambda}^{(i)}({\bf x}_{\perp,i})
\bar{\psi}^{(i)}_{\lambda}({\bf x}^\prime_{\perp,i})\, .
\label{G^2}
\end{equation}
This equation illustrates that for large imaginary times the integrand in 
Eq.\ (\ref{GE}) decays as $\exp[-E_Bt]$. Notice that this representation 
is valid for $E_B>0$. By using 
\begin{equation}
\frac{\mu}{2\pi r}=
\int_{0}^{\infty}dt\,\Big(\frac{\mu}{2\pi t}\Big)^{3/2}
e^{-r^2\mu/2t}\, 
\end{equation}
we find  
\begin{equation}
\zeta_{E}({\bf R}_\perp,{\bf R}_\perp^\prime) 
=\int_0^{\infty}\frac{dt}{2\mu}\,\Big[e^{Et}
G_t({\bf R}_\perp,0;{\bf R}_\perp^\prime,0)
-\left(\frac{\mu}{2\pi t}\right)^{3/2}
\delta({\bf R}_\perp-{\bf R}_\perp^\prime)\Big]\, .
\label{kernel2}
\end{equation}
To show that the integral in Eq.\ (\ref{kernel2}) converges 
also for small $t$, we expand $G_t({\bf R}_\perp,0;{\bf
R}_\perp^\prime,0)$ with respect to $t$,  
see Appendix A. We find 
\begin{eqnarray}
\lim_{t\rightarrow0}G_t({\bf R}_\perp,0;{\bf R}_\perp^\prime,0) & =
& \left(\frac{\mu}{2\pi t}\right)^{3/2}
\delta({\bf R}_\perp-{\bf R}_\perp^\prime)
-t^{-1/2} \left(\frac{\mu}{2\pi}\right)^{3/2}\nonumber\\
& & \times\left[\frac{{\bf P}^2_\perp}{2M}+V_1({\bf R}_\perp)+V_2({\bf
R}_\perp)\right] \, . 
\label{smallt}\end{eqnarray}
Thus $\zeta_E$ can be regarded as a regular
operator acting on the space ${\cal L}^2$ of  square-integrable
functions.
We note in passing that if the two single-particle transverse Hamiltonians 
$H_{\perp,i}$ commute with the angular momentum operators $L_{z}$, then
also $\zeta_E$ commutes 
with $L_{z}$. This follows by observing that in this case we can choose 
for the eigenbasis $\{\psi^{(i)}_\lambda\}$ a set of eigenvectors of
$L_{z}$, and the product of two eigenvectors of $L_{z}$ is
 still an eigenvector of $L_{z}$. 
Hence  $\zeta_E({\bf R}_\perp,{\bf R}_\perp^\prime)$ can be written
as a sum of 
projectors onto states with definite angular momentum.
A similar conclusion can be drawn regarding parity
 symmetry, when considering non-cylindrical confining potentials that
obey this symmetry. 

Using the transverse
 non-interacting ground state,
\begin{equation}
\psi_0({\bf R}_\perp,{\bf r}_\perp)=
\psi^{(1)}_{0} \left({\bf R}_\perp+\frac{\mu}{m_1}{\bf r}_\perp\right)
 \psi^{(2)}_{0}\left({\bf R}_\perp-\frac{\mu}{m_2}{\bf r}_\perp\right)
 \, , 
\label{tgt}
\end{equation}
provided the overlap integral
$\int d{\bf R}^{\prime}_\perp\bar{\psi}_{0}({\bf R}_\perp^\prime,0)
 f({\bf R}^{\prime}_\perp)\neq0$,
 the integrand in Eq.\  (\ref{Gensol}) decays as $\exp[-E_Bt]$. This
 defines a spatial scale $a_B$ for the longitudinal size of the corresponding
 bound state, 
$a_B=1/\sqrt{\mu E_B}.$
For large $E_B$, $a_B$ is small and we have very
tight pairs. This constitutes the {\it dimer limit\/}. On the other
hand,  for small $E_B$, atom pairs are very elongated. This regime is termed
 {\it BCS limit}. In the following, we investigate
both limits in greater detail. 

\subsection{Dimer limit}  

For large binding energies, the atom-atom interaction dominates
over the confinement.
Due to the exponential factors in Eq.\ (\ref{kernel2}), only small imaginary 
times contribute significantly 
to the integral, and we can substitute $G_t$ with the short-time expansion
 (\ref{smallt}) as derived in the Appendix A, yielding
\begin{eqnarray}
\zeta_{E}({\bf R}_\perp,{\bf R}_\perp^\prime) & \simeq &
\left(\frac{\mu}{2\pi }\right)^{3/2}\int_{0}^{\infty}\frac{dt}{2\mu}\,
\left(  t^{-3/2}
\left(e^{Et}-1\right)
\delta({\bf R}_\perp-{\bf R}^{\prime}_\perp) \right. \nonumber \\
& &  \left. -e^{Et}t^{-1/2}
\langle {\bf R}_\perp | H_\perp|{\bf R}_\perp' \rangle 
\right)  \, .
\end{eqnarray}
Hence, the operator $\zeta_{E}$ now shares 
 eigenfunctions with  
$H_\perp = {\bf P}^2_\perp/2M+V_1({\bf R}_\perp)+V_2({\bf R}_\perp)$. 
For (identical) parabolic confinement potentials,
$H_\perp$ is exactly  the decoupled COM Hamiltonian.
Let us denote  the 
eigenfunctions and eigenenergies of $H_\perp$ as 
$\phi_\lambda({\bf R}_\perp)$ and $E^{(\phi)}_\lambda$, respectively.
 Substituting $\phi_\lambda({\bf R}_\perp)$ into Eq.\ (\ref{eigeq})   yields
after some algebra
\begin{equation}
-\frac{1}{4\pi a}=-\frac{\sqrt{2\mu
|E|}}{4\pi}\left(1+\frac{E^{(\phi)}_\lambda}{2|E|}\right)\simeq-
\frac{\sqrt{2\mu E_B}}{4\pi}\, . \label{mindy}
\end{equation}
In the second relation, we have used Eq.\  (\ref{E_B}). 
{}From this, we directly obtain the binding energy in the 
dimer limit $a\to 0^{+}$ as 
\begin{equation}
E_B\approx\frac{1}{2\mu  a^2}
\,,  \label{Dimer}
\end{equation}
which coincides with the result obtained in free (3D) space without confinement. 
\subsection{BCS limit} 
The scattering channel with lowest energy, corresponding to the
 transverse non-interacting ground state $\psi_0$, opens at the
 energy threshold $E=E_0$. 
For $E_B\rightarrow0^+$, as the energy  approaches this threshold,
 the  term with $\lambda_1=\lambda_2=0$ dominates
in Eq.\  (\ref{G^2}),  and  yields 
 in Eq.\ (\ref{kernel2}) the contribution
\begin{equation}
\sqrt{\frac{1}{8\mu E_B}}
\psi_{0}({\bf R}_\perp,0)
\bar{\psi}_{0}({\bf R}_\perp^\prime,0)
\,  ,
\label{1stch}
\end{equation}
which diverges for $E_B\rightarrow0^+$.  
All other channels are still closed at $E=E_0$ and give finite
contributions in Eq.\ (\ref{G^2}). 
This observation suggests a useful separation of the total Hilbert
space into a part   ${\cal H}_{\rm o}$ corresponding to the open channel
(or lowest-energy scattering channel) 
and a part ${\cal H}_{\rm e}$ perpendicular to that. With this
separation, terms yielding a finite contribution at $E_B\to 0^+$
can be summarized in the Green's function
\begin{equation}
\tilde{G}_t({\bf R}_\perp,{\bf r};{\bf R}^\prime_\perp,0)=
\langle{\bf R}_\perp,{\bf
r}|\exp[- \tilde{H}_0t]|{\bf R}^\prime_\perp,0\rangle \, , 
\end{equation}
where $\tilde{H}_0$ is the projection of 
 $H_0$ onto the Hilbert subspace ${\cal H}_{e}$.
 We then define a new integral kernel,
\begin{equation}
\tilde{\zeta}_{E}({\bf R}_\perp,{\bf R}_\perp^\prime) 
=\int_0^{\infty}\frac{dt}{2\mu}\,\Big[e^{Et}
\tilde{G}_t({\bf R}_\perp,0;{\bf R}_\perp^\prime,0)
-\left(\frac{\mu}{2\pi t}\right)^{3/2}
\delta({\bf R}_\perp-{\bf R}_\perp^\prime)\Big]\,,
\label{kernel3}
\end{equation}
which is also well-defined for energies above the threshold $E=E_0$.

For small $E_B$, Eq.\ (\ref{eigeq}) is most conveniently solved by 
expanding $f({\bf R}_\perp)$  in an
 orthonormal basis $|j\,\rangle$ according to 
\begin{equation}\label{deff}
|\,f\,\rangle = \sum_j f_j |\,j\,\rangle, \quad
 f_j = \int d{\bf R}_\perp \langle\, j\,|{\bf R}_\perp
\rangle f({\bf R}_\perp),
\end{equation}
where the basis state $|\,0\,\rangle$ corresponds to
\begin{equation}\langle {\bf R}_\perp|\,0\,\rangle= 
c\psi_{0}({\bf R}_\perp,0),
 \label{0}\end{equation} 
with normalization constant $c$.  Although
$\psi_0({\bf R}_\perp,{\bf r})$ 
is a normalized element of the two-particle 
Hilbert space, this 
 does not imply that $\psi_0({\bf R}_\perp,0)$ 
is an element of the COM Hilbert space with norm unity. 
In fact, the normalization constant $c$ has to be computed explicitly 
and generally depends on the particular confinement.  
In this basis,   Eq.\  (\ref{eigeq}) assumes the compact form
\begin{equation} \label{eigeq3}
-\frac{|f\rangle}{4\pi a}=\zeta_E |f\rangle=\left( \sqrt{\frac{1}{8\mu
E_B}}
\frac{|0\rangle\langle0|}{c^2}
+\tilde{\zeta}_E\right) |f\rangle.
\end{equation}
 $|0\rangle$ is an approximate eigenstate for small $E_B$, since all  the 
matrix elements are finite apart from $\langle0|\zeta_E|0\rangle$ which 
diverges according to  
\begin{equation}
\langle0|\zeta_{E}|0\rangle\simeq\sqrt{\frac{1}{8\mu
E_B}} \frac{1}{c^2}+
\langle0|\tilde{\zeta}_E|0\rangle \, .
\label{martin}
\end{equation}
Substituting this in Eq.\ (\ref{eigeq3}) yields
\begin{equation}
-\frac{1}{4\pi a}\simeq \sqrt{\frac{1}{8\mu
E_B}} \frac{1}{c^2}+
\langle0|\tilde{\zeta}_E|0\rangle \, .
\end{equation}
Neglecting the last term, the relation for the binding energy $E_B$ is solved 
in  the BCS limit $a\to 0^-$,
\begin{equation}
E_B\approx \frac{2a^2\pi^2}{\mu c^4} \, .  \label{BCS}
\end{equation}
%
%
\section{Scattering solutions\label{sec.scatter}}
In this section, 
 we focus on scattering solutions at low energies $E$ slightly 
above $E_0$, where exactly one transverse channel is open. 
 Then the incoming state is given by 
\begin{equation}
\Psi_0=e^{ikz}
\psi_{0}^{(1)}({\bf x}_{1,\perp})
\psi_{0}^{(2)}({\bf x}_{2,\perp}) \, ,
\end{equation} 
which describes
two incoming atoms with (small) relative longitudinal momentum 
$k=\sqrt{2m(E-E_0)}$ in the (transverse) single-particle ground states
$\psi_{0}^{(1)}$ and $\psi_{0}^{(2)}$, respectively. 
%
\subsection{One-dimensional scattering length $a_{\rm 1D}$}

As done in Sec.~\ref{sec.boundstates}, 
we split off the contribution from the 
open channel, 
\begin{equation}
\fl G_E({\bf R}_\perp,{\bf r};{\bf R}^\prime_\perp,0)  = 
\psi_{0}({\bf R}_{\perp},{\bf r}_{\perp})
\bar{\psi}_{0}({\bf R}_{\perp}^\prime,0)\frac{i\mu}{k}e^{ik|z|}
 +\int_0^\infty dt\,e^{Et}
\tilde{G}_t({\bf R}_\perp,{\bf r};{\bf R}_\perp^\prime,0)\, ,
\label{barnes} 
\end{equation}
where $\tilde{G}_t({\bf R}_\perp,z;{\bf R}_\perp^\prime,0)$ is the
Green's function restricted to ${\cal H}_{e}$,
which is well-defined also above $E_0$. 
Inserting Eq.~(\ref{barnes}) into Eq.~(\ref{Gensol})
yields for $|z|\rightarrow\infty$ the standard
scattering solution,
\begin{equation}
\Psi({\bf R},{\bf r})=\psi_0({\bf R}_\perp,{\bf r}_\perp)
(e^{ikz}+f_e(k)e^{ik|z|}),
\end{equation}
with scattering amplitude
\begin{equation}\label{sa}
f_e(k)=\frac{i}{2k}\int d{\bf R}^\prime_\perp 
\bar{\psi}_0({\bf R}_\perp^\prime,0)
f({\bf R}_\perp^\prime)\, ,
\end{equation}
whereas for short distances,  also the term  
$\int d{\bf R}^\prime_\perp\int_0^\infty dt\,e^{Et}
\tilde{G}_t({\bf R}_\perp,{\bf r};{\bf R}_\perp^\prime,0)
f({\bf R}_\perp^\prime)$
appears in the scattering solution. 
Since the energy is well below the continuum threshold for the closed 
channels, this must be regarded as a sum over localized states. 
Enforcing the boundary condition (\ref{bc}) 
 then leads to an integral equation for $f({\bf R}_\perp)$,
\begin{eqnarray} \label{gastone} 
 -\frac{f({\bf R}_\perp)}{4\pi a} & = &  
 \int d{\bf R}_\perp^\prime\tilde{\zeta}_E({\bf R}_\perp,{\bf R}_\perp^\prime)
f ({\bf R}_\perp^\prime)
 \nonumber \\ 
& &+\psi_0({\bf R}_\perp,0)
 +\frac{i\psi_0({\bf R}_\perp,0)}{2k} \int d{\bf R}_\perp^\prime
\bar{\psi}_0({\bf R}_\perp^\prime,0) f({\bf R}_\perp^\prime) \, .
\end{eqnarray}
This integral equation is most conveniently solved by 
again
expanding $f({\bf R}_\perp)$ in the orthonormal basis $\{|j\rangle\}$ 
introduced in the previous section. 
Thereby, we can express Eq.~(\ref{gastone}) in compact notation, 
\begin{equation} \label{ham}
-\frac{|f\rangle}{4\pi a}=\frac{|0\rangle}{ c}+\frac{i}{2k}
\frac{|0\rangle}{c^2}\langle0|f\rangle+\tilde{\zeta}_E |f\rangle,
\end{equation}
which is formally solved by 
\begin{equation}\label{solution}
|f\rangle= \frac{-1/c}{1-i/(k a_{\rm 1D})}
\left(\tilde{\zeta}_E+\frac{1}{4\pi a}\right)^{-1}|0\rangle.
\end{equation}
The parameter $a_{\rm 1D}$ follows in the form
\begin{equation}\label{cl}
a_{\rm 1D}=-\frac{2c^2}{\langle 0| [\tilde{\zeta}_{E}+1/(4\pi
a)]^{-1}|0\rangle}\, .
\end{equation}
{}From Eq.\ (\ref{sa}), $f_e(k)=-1/(1+ik a_{\rm 1D})$,
which allows to identify 
$a_{\rm 1D}$ with the {\it 1D scattering length\/}. 
Having introduced this parameter, the 1D atom-atom interaction
potential can then be written in an effective form according to 
\begin{equation}
V_{\rm 1D}(z,z')
=g_{\rm 1D} \delta(z-z') \, ,
\end{equation}
 with interaction strength 
  $g_{\rm 1D}=-1/(\mu a_{\rm 1D})$  \cite{olshanii}.  
For very low energies, $k\to0$, we can now formally set  $E=E_0$ in 
Eq.\  (\ref{cl}). For a confining  trap, $\tilde{\zeta}_{E_0}$ is an Hermitian operator 
with discrete spectrum $\{\lambda_n\}$ and eigenvectors $|e_n\rangle$,
which eventually have to be determined for the particular 
Hamiltonian. Thus we find
\begin{equation}\label{u0}
g_{\rm 1D}= \frac{1}{2 \mu c^2}
 \sum_n \frac{|\langle 0|e_n \rangle|^2}{\lambda_n+1/(4\pi a)}.
\end{equation}
 This result has interesting consequences for the two-body
 interaction. The denominator can become singular for particular
 values of $a$, thereby generating a CIR. 
 Every eigenvalue $\lambda_n$ corresponds to a 
different CIR, unless the overlap $\langle 0|e_n\rangle$
vanishes due to some underlying symmetry of the Hamiltonian. 
We anticipate that for identical parabolic confinement potentials,
 the decoupling of the
COM motion implies that only one resonance is permitted. 
 For confining potentials with cylindrical 
symmetry,  there is a resonance for each
 eigenvector of $\tilde{\zeta}_{E_0}$ with zero angular momentum. For
 confining potentials obeying parity symmetry, the eigenstates 
  $|e_n\rangle$ must be even. 
These two symmetries allow in principle for infinitely many resonances. 
In practice, however, only few of them can be resolved because
the resonances become increasingly sharper 
 when $|\langle 0|e_n \rangle|^2\to 0$, making them
difficult to  detect.

\subsection{Interpretation of the CIR as Feshbach resonances}

A very simple and illuminating
analysis,  similar to that for standard Feshbach resonances \cite{timmer},
is also possible for
the CIR. The two-particle Schr\"odinger equation 
can be rewritten as an  effective Schr\"odinger equation for the 
scattering states in the open channel, $(E-H_{\rm
eff}){\cal P}|\Psi\rangle=0$, with 
the effective Hamiltonian 
\begin{equation}
H_{\rm eff}=H_{\rm open}+{\cal P} H{\cal M}\frac{1}{E-H_{\rm closed}}{\cal M}H{\cal P} \, .
\end{equation}
Here,  $H_{\rm open}={\cal P}H{\cal P}$ and $H_{\rm closed}={\cal M}H{\cal M}$, 
where ${\cal P}$ and ${\cal M}$
are projectors to open  and closed channels, respectively.
  This equation can be expressed in terms of the closed-channel eigenstates
  $|\Phi_n\rangle$,
\begin{equation}
H_{\rm eff}=H_{\rm open}+{\cal P}H\sum_{n}
\frac{|\Phi_n\rangle\langle\Phi_n|}{E-E_n}H{\cal P} \, , 
\end{equation} 
with $H_{\rm closed}|\Phi_n\rangle=E_n|\Phi_n\rangle$.
This implies that a Feshbach-like resonance is possible 
at zero momentum if two conditions are
 fulfilled. First, there exists a solution of 
 $(E_0-H_{\rm closed})|\Phi\rangle=0$, i.e.,
 $|\Phi\rangle$ is a 
bound state  of $H_{\rm closed}$  with energy 
$E=E_0$.
 Second, $|\Phi\rangle$ must be coupled to the open
 channel, ${\cal P}H|\Phi\rangle\neq0$.

Within the pseudopotential approximation, the equation  
$(E_0-H_{\rm closed})|\Phi\rangle=0$ 
 is solved in terms of the Green's function
\begin{equation}
{\cal M}G_{E_0}({\bf R}_\perp,{\bf r};{\bf R}^{\prime}_\perp,0){\cal M}
=\int_0^\infty dt\,e^{E_0t}
\tilde{G}_t({\bf R}_\perp,{\bf r};{\bf R}_\perp^\prime,0)\,
\end{equation}
by the state 
\begin{equation}
\Phi({\bf R}_\perp,{\bf r})=
\int d{\bf R}^{\prime}_\perp\,
{\cal M}G_{E_0}({\bf R}_\perp,{\bf r};{\bf R}^{\prime}_\perp,0){\cal M}
\frac{f({\bf R}^{\prime}_\perp)}{2\mu}\, , \label{Gensol2}
\end{equation}
together with boundary condition
\begin{equation}
\Phi({\bf R}_\perp,{\bf r}\to 0) \simeq 
\frac{f({\bf R}_\perp)}{4\pi r}\left(1-\frac{r}{a}\right)
\label{bc2}\, . 
\end{equation}
This leads to the eigenvalue equation 
\begin{equation}
-\frac{|f\rangle}{4\pi a}=\tilde{\zeta}_{E_0} |f\rangle \, ,
\end{equation}
which is solved by the eigenvectors $|e_n\rangle$
introduced above. This yields  
 $a=-1/(4\pi \lambda_n)$, implying that   there is a
 bound state $|\Phi\rangle$ of $H_{\rm closed}$  with energy   equal to the
energy of the incoming wave, corresponding to the resonances found in the 
previous subsection. The CIR is then in complete
analogy to a zero-momentum Feshbach resonance. Due to the small but finite
 coupling to the closed channels, two incoming particles initially in the open
channel  visit the closed channels  during the
 scattering process. This process is strongly intensified when 
 a bound-state exists whose energy is close to 
 the continuum threshold. Then, a scattering resonance results.
 Note that such a bound state can be occupied only virtually by two particles 
during the scattering process. Hence  from now on we will refer to such a 
bound state as a virtual bound state.

It is also possible to recover the overlap 
condition $\langle 0|e_n\rangle\neq0$ in this framework. 
In fact, 
\begin{equation}
\fl {\cal P}H\Phi({\bf R}_\perp,{\bf r})=
{\cal P}V({\bf r})\Phi({\bf R}_\perp,{\bf r}) 
=-{\cal P}\frac{\langle{\bf R}_\perp|e_n\rangle}{2\mu} \delta({\bf r}) 
=-\psi_0({\bf R}_\perp,{\bf r}_\perp)\delta(z) 
\frac{\langle 0|e_n\rangle}{2\mu c}\, ,
\end{equation}
since $|\Phi\rangle$ fulfills Eq.~(\ref{bc2}) with 
$f({\bf R})=\langle{\bf R}_\perp|e_n\rangle$.
Hence, the two overlap conditions
\begin{equation}
{\cal P}H|\Phi\rangle\neq0 \qquad 
\Leftrightarrow \qquad \langle 0|e_n\rangle\neq0\,
\label{overlap}
\end{equation} 
are equivalent. When they
are not fulfilled, there exists a virtual  bound state with energy $E_0$,
 but it is  not coupled to  the incoming wave.

\section{Special case of harmonic confinement\label{sec.parabolic}}

In the previous sections, we have formulated the theory for a general
confining potential and for two different atomic species. 
As a simple illustration, we now consider the case of harmonic confinement,
$V_{i}({\bf x}_i)=
m_{i}\omega^2_{i}{\bf x}_{i\perp}^2/2$. In COM and relative
coordinates,
\begin{eqnarray}
V_{\rm conf}({\bf R}_\perp,{\bf r}_\perp)
& = & \frac{1}{2}\left(m_{1}\omega^2_{1}+m_{2}\omega^2_{2}\right)|{\bf R}_\perp|^2
+\frac{1}{2}
\left(\frac{\mu^2}{m_1}\omega^2_{1}+\frac{\mu^2}{m_2}\omega^2_{2}\right)
|{\bf r}_\perp|^2 \nonumber \\ 
& & +\mu\left(\omega^2_{1}-\omega^2_{2}\right)
{\bf r}_\perp\cdot{\bf R}_\perp \, .
\end{eqnarray}
In general, the COM and the relative coordinate do not decouple, and in order
 to find the scattering and bound-state solutions, we have to follow the
 procedure outlined in the previous sections. 
 To that end, we label the single-particle 
transverse states  by quantum numbers $\lambda=\{m,n\}$, where 
$m$ is the integer angular momentum  and $n$  the
 integer radial quantum number.
The eigenenergies   and  -states of the 2D
 harmonic oscillator
\[
E^{(i)}_\lambda=\omega_{i}\epsilon_{n,m},\quad
\psi^{(i)}_\lambda=\frac{1}{a_i}\psi_{n,m}\left(\frac{{\bf
x}_{\perp}}{a_i}\right) \, , 
\]
with the oscillator lengths $a_i=(m_i\omega_i)^{-1/2}, i=1,2$, can be
expressed in terms of the quantities
\[
\epsilon_{n,m}=2n+|m|+1\qquad {\rm and } \qquad
\psi_{n,m}({\bf x}_{\perp})=e^{im\phi}R_{n,m}(|{\bf x}_{\perp}|),
\]
where
\[
R_{n,m}(\rho)=\frac{1}{\sqrt{\pi}}\left(\frac{n!}{(n+|m|)!}\right)^{1/2}
e^{-\rho/2}\rho^{|m|}L^{|m|}_n(\rho^2),
\]
with $L^{|m|}_n(x)$ being  the  standard Laguerre polynomials.
 A convenient choice for the orthonormal basis 
$|j\rangle$  introduced in Eq.\ (\ref{deff}) is then given by
\begin{equation}
\langle{\bf R}_\perp|j\rangle =\langle{\bf R}_\perp|m,n\rangle
=\frac{1}{a_{M}}\psi_{n,m}\left(\frac{|{\bf
R}_{\perp}|}{a_{M}}\right) \, , 
\label{lagpol}
\end{equation}
with the length scale $a_{M} =(m_1\omega_1+m_2\omega_2)^{-1/2}$.
In particular, we find for $|0\rangle = |0,0\rangle$ that 
$\langle{\bf R}_\perp|0\rangle$ fulfills Eq.\ (\ref{0}) with 
$c=\sqrt{\pi}a_1a_2/a_{M}$. 

The single-particle imaginary-time propagator for  a 2D harmonic
oscillator with length scale $a_0$ and frequency $\omega$ is given by
\begin{eqnarray}
\fl\sum_{\lambda}e^{-\omega\epsilon_\lambda t}\frac{1}{a_{0}^2}
\psi_\lambda\left(\frac{{\bf x}_\perp}{a_0}\right)
\bar{\psi}_\lambda\left(\frac{{\bf x}_\perp^\prime}{a_0}\right)
=\frac{1}{\pi a_{0}^2}
\frac{e^{-\omega t}}{1-e^{-2\omega t}}
\exp{\left[-
\frac{{\bf x}_{\perp}^2+{\bf x}_{\perp}^{\prime2}}{2a_0^2}\coth(\omega t)
+\frac{{\bf x}_{\perp}\cdot{\bf x}_{\perp}^\prime}{a_0^2\sinh(\omega t)}
\right]}.  \nonumber \\
\label{sum}
\end{eqnarray}
Inserting this into Eq.\ (\ref{G^2}) with
 ${\bf x}_{\perp,i}={\bf R}_\perp$,
 ${\bf x}_{\perp,i}^\prime={\bf R}_\perp^\prime$ and $z=0$,
we find
\begin{eqnarray}
G_t({\bf R}_\perp,0;{\bf R}^\prime_\perp,0) & = & 
\sqrt{\frac{\mu}{2\pi t}}\frac{\beta(1-\beta)}{\pi^2 a_{M}^4}
\frac{e^{-\omega_1 t}}{1-e^{-2\omega_1 t}}
\frac{e^{-\omega_2 t}}{1-e^{-2\omega_2 t}}\nonumber \\ 
& & \times \exp{\left[-
\frac{{\bf R}_{\perp}^2+{\bf R}_{\perp}^{\prime2}}{2a_{M}^2}f(t)+
\frac{{\bf R}_{\perp}\cdot{\bf R}_{\perp}^\prime}{a_{M}^2}g(t)
\right]},
\label{G^2b}
\end{eqnarray} 
where we have introduced $\beta=a_{M}^2/a_1^2$ and 
\begin{eqnarray} \label{fgfunc}
f(t)&=&\beta\coth(\omega_1 t)+(1-\beta)\coth(\omega_2 t)\,, \nonumber \\
g(t)&=&\beta\sinh^{-1}(\omega_1 t)+(1-\beta)\sinh^{-1}(\omega_2 t)\, .
\end{eqnarray}
 In order to compute explicitly the operators $\zeta_E$ and
 $\tilde{\zeta}_E$, 
 we still have to
 project onto the discrete basis  $\{|j\rangle\}$ and to perform the 
imaginary-time  integral for each matrix element. In general, this 
cannot  be achieved analytically, and one has to
 resort to a  numerical evaluation.
Only for  $\omega_1=\omega_2$, a complete
 analytical solution is possible. Since 
 the COM degrees of freedom separate,
 this solution is a trivial extension of Ref.\  \cite{olshanii}. 
 Nonetheless, along 
with the general analysis of the previous section, 
it provides a 
 physical picture for weak interaction between the COM and
 the relative degrees of freedom. 
 \subsection{Identical frequencies}
For $\omega_1=\omega_2=\omega$, the COM 
and relative coordinates separate, $H=H_{\rm rel}+H_{\rm COM}$,
with 
\begin{equation}
H_{\rm rel}=\frac{{\bf p}^2}{2\mu}+\frac{1}{2}\mu\omega^2{\bf
r}_\perp^2+V({\bf r})\, , \quad
H_{\rm COM}=\frac{{\bf P}^2}{2M}+ \frac{1}{2}M\omega^2{\bf R}_\perp^2 \, .
\end{equation}
In this case, we can consider the two-particle system being (asymptotically)
 in the 
ground state of the decoupled COM Hamiltonian, and just solve 
the relative problem \cite{olshanii,bergeman}.
Moreover, with $f(t)=\coth(\omega t)$ and $g(t)=\sinh^{-1}(\omega t)$,
the Green's function (\ref{G^2b}) simplifies to 
\begin{equation}
\fl G_t({\bf R}_\perp,0;{\bf R}^\prime_\perp,0)=
\sqrt{\frac{\mu}{2\pi t}}\frac{\beta(1-\beta)}{\pi a_{M}^2}
\frac{e^{-\omega t}}{1-e^{-2\omega t}}
\sum_{n,m}e^{-\omega\epsilon_{n,m} t}\frac{1}{a_{M}^2}
\psi_{n,m}\left(\frac{{\bf R}_\perp}{a_{M}}\right)
\bar{\psi}_{n,m}\left(\frac{{\bf R}_\perp^\prime}{a_{M}}\right) \, .
\label{G^2c}
\end{equation}
In this case, $|n,m\rangle$ is an 
eigenstate of the decoupled Hamiltonian $H_{\rm COM}$, and describes
 the COM motion also for finite ${\bf r}$.
Moreover, $a_{M}=(M\omega)^{-1/2}$ and
$a_\mu=a_{M}/(\beta(1-\beta))=(\mu\omega)^{-1/2}$ are the characteristic 
lengths associated 
with $H_{\rm COM}$ and $H_{\rm rel}$, respectively. 
Inserting Eq.\ (\ref{G^2c}) into Eq.\ (\ref{kernel2}) and 
rescaling $t$ by $2\omega$, we obtain 
\begin{equation} 
\zeta_E=\sum_{n,m}\frac{|n,m\rangle\langle n,m|}{4\pi a_\mu}
\int_0^\infty\,\frac{dt}{(\pi t)^{1/2}}
\left(\frac{e^{-\Omega_{n,m}(E) t}}{1-e^{-t}}-\frac{1}{t}\right),
\label{zetaw}
\end{equation}
with $\Omega_{n,m}(E)=(1+\epsilon_{n,m}-E/\omega)/2$.  The integral on
the rhs of Eq.\  (\ref{zetaw}) is related to the 
integral representation of the Hurvitz zeta 
function  $\zeta(1/2,\Omega_{n,m})$ \cite{Mora05,Grad}. 

\subsubsection{Bound states} 

\begin{figure}[h]
\begin{center}
\epsfig{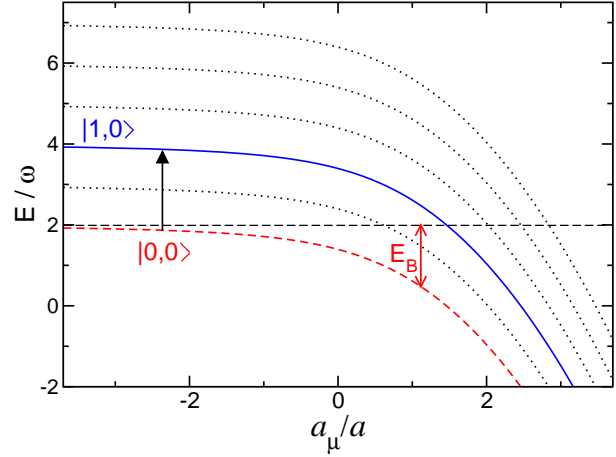}
\caption{Bound state energies $E$  
 as a function of  $a_\mu/a$ for harmonic confinement with equal frequency
$\omega$ for both atoms. The dashed red curve indicates the bound
state energy of the ground state $|0,0\rangle$. Its binding energy
$E_B$ is given by the distance to the horizontal dashed line 
indicating the continuum threshold  for the open channel.  
  The blue curve marks 
the bound state energy of the virtual bound state relevant for the 
low-energy scattering. It is obtained by a vertical
shift of the ground-state energy by $2\omega$, and
coincides with the three-fold degenerate bound-state energy indicated
as solid curve. 
The black dotted curves give the bound-state energies  of the 
excited transverse
states, obtained by a vertical shift of the ground-state result. 
 \label{fig.1}}
\end{center}
\end{figure}

The condition given in Eq.\ (\ref{eigeq}) 
for a bound state with transverse configuration $|{n,m}\rangle$
 translates into
\begin{equation}
\zeta\left(\frac{1}{2},\Omega_{n,m}\right)=-\frac{a_\mu}{a}\,.
\label{Hz}
\end{equation}
The zeta function is monotonic, and has the asymptotic 
scaling behavior
\begin{equation}
\zeta\left(\frac{1}{2},\Omega \ll 1\right) \approx \Omega^{-1/2}, 
\quad \zeta\left(\frac{1}{2},\Omega\gg 1\right ) \approx -2\sqrt{\Omega}\, . 
\label{ska}
\end{equation}
Inverting Eq.\ (\ref{Hz}), we recover the bound-state energy 
found in Ref.~\cite{bergeman}. The corresponding
result is plotted in Fig.\ \ref{fig.1}. 
As  an immediate consequence of the decoupling
 of the COM degrees of freedom, the 
$\epsilon_\lambda$-fold degenerate
energies corresponding to excited transverse 
configurations follow from the COM transverse ground state by
a shift along the ordinate in steps of $\omega$. This is indicated by
the dotted curves  in Fig.\ \ref{fig.1}. 
Notice that for energies above $E_0=2\omega$, corresponding to
$E_B=2\omega\Omega_{0,0}(E)<0$, there exists an open channel, but the 
solutions associated with  COM excited states are orthogonal to
 it. For this reason, the relevant condition for a bound state to exist
 with 
transverse configuration $|{n,m}\rangle$ is $\Omega_{n,m}(E)>0$.      
>From the scaling behaviors in Eq.\ (\ref{ska}), we find the limiting
behaviors of the energy of the bound state at $|a_\mu/a|\ll 1$ as
\begin{eqnarray}
E_{B, {n,m}} &\approx& \frac{ 1}{2\mu a^2 }
\quad {\rm for} \quad a>0 \, ,\nonumber \\
E_{B, {n,m}} &\approx& \frac{ 2 a^2}{\mu a_\mu^4 }
\quad {\rm for} \quad a<0\, , 
\end{eqnarray} 
 see Eqs.\ (\ref{Dimer})  and  (\ref{BCS}),
with $c=\sqrt{\pi}a_\mu$ and $E_{B, {n,m}}=\omega \Omega_{n,m}$. Hence, in this 
highly degenerate case,
  there is exactly one bound state for each transverse configuration and 
each scattering length $a$. 

\subsubsection{Scattering states} 

In order to identify resonant bound states of the closed channel,
 and the corresponding
zero-momentum CIR, we subtract the
contribution of the lowest-energy scattering channel
in Eq.\ (\ref{zetaw}), and obtain 
\begin{eqnarray} \nonumber
\tilde{\zeta}_E &=& \zeta_E-\frac{|0,0\rangle\langle 0,0|}{4\pi a_\mu}
\int_0^\infty\frac{dt}{(\pi t)^{1/2}}e^{-\Omega_0(E)t }\\
&=& \sum_{{n,m}}\frac{|{n,m}\rangle\langle {n,m}|}{4\pi a_\mu}
\zeta(1/2,\tilde{\Omega}_{n,m}(E)),
\label{Zetapar}
\end{eqnarray} 
with $\tilde{\Omega}_{0,0}(E)=\Omega_{0,0}(E)+1$ and 
$\tilde{\Omega}_{n,m}(E)=\Omega_{n,m}(E)$
 for $n+|m|>0$. Hence the curve corresponding to the COM ground state is 
shifted vertically by $2\omega $ and coincides with the curve 
corresponding to the excited states  $|1,0\rangle$, $|0,2\rangle$ and 
$|0,-2\rangle$. Moreover, the coupling condition in Eq.\ (\ref{overlap}) 
becomes $\langle0,0|n,m\rangle\neq0$, and is fulfilled only for
$n=m=0$. 
Though there  are in principle 
infinitely many closed-channel bound states with energy 
$2\omega$ (one for each curve),  only one scattering
resonance exists,  
since only one of them is coupled to the incoming scattering wave.
Inserting Eq.\ (\ref{Zetapar}) into
 Eq.\ (\ref{u0}) we recover for the 1D interaction strength $g_{\rm 1D}$ the
 well known result \cite{olshanii}
\begin{equation}
g_{\rm 1D}=2\omega a_\mu\left(\frac{a_\mu}{a}-{\cal C}\right)^{-1}.
\end{equation}

\subsubsection{Physical picture for the weakly interacting case 
\label{remarks}} 
 
 When $\omega_1\neq\omega_2$ but $\omega_1\approx \omega_2$, 
  a weak coupling to the COM degrees of freedom is generated, with two
  important consequences: 
 (i) the degeneracies of the bound-state energies are lifted, and 
(ii) the coupling to the 
other higher-lying bound states is non-zero. 
Since the operators $\tilde{\zeta}_E$ and ${\zeta}_E$  commute with the
$z$-component $L_z$ of the angular momentum,  
 the bound states are still labeled by the quantum numbers $\{n,m\}$. 
 As far as the scattering solutions are concerned, the
 incoming wave is coupled only to states with angular momentum quantum 
number $m=0$. Since
$\langle0,0|\tilde{\zeta}_E|0,0\rangle\approx\langle1,0|\tilde{\zeta}_E|1,0\rangle$, 
a small off-diagonal element $\langle0,0|\tilde{\zeta}_E|1,0\rangle$ 
is sufficient to couple the bound state with $\{n,m\}=\{1,0\}$ to the
incoming wave, yielding an additional CIR.  
  As far as bound states are concerned, 
solutions with $E>E_0$ and $m=0$ leak into the open channel, and cannot be
regarded as 
 localized bound states. Hence, for $|a_\mu/a|\gg 1$ and $a<0$,  
there is only one bound state with zero angular momentum. 
In the opposite dimer limit,  however, we encounter many 
dimer bound states.

\subsection{The case $\omega_1\ne \omega_2$: Relation to
experiments}

The case   $\omega_1\ne \omega_2$ is relevant for
experiments 
involving two different atom species trapped in magnetic or  optical traps
 \cite{Exp1,Exp2,Exp3}. 
 For instance, in optical traps the confining potential depends on the
 detuning 
$\Delta=\omega_{\rm las}-hc/\lambda$ of the laser frequency $\omega_{\rm las}$
 from the characteristic frequency $hc/\lambda$ associated with the optical
 transition $ns\rightarrow np$, and is therefore different for two different
 atom species. This conclusion also applies to magnetic traps 
if the atoms are confined
 in hyperfine states with  different projection of the magnetic moment along 
the magnetic field.
 As a concrete example, let us consider a mixture of Bosonic $^{87}$Rb atoms 
and Fermionic $^{40}$K atoms. Sympathetic 
cooling  has allowed to create  an ultracold
 mixture of these two elements. By loading such a gas into a dipole trap and 
 sweeping an external magnetic field, it has been possible \cite{Exp3} 
 to identify
 three heteronuclear Feshbach resonances and to measure the 3D
 interspecies scattering length $a=-14$ nm. It seems  feasible to tune
 the magnetic field near a Feshbach resonance and to observe the interspecies 
CIR. It is hence very interesting to know how many of
them can be expected and to study their locations.  

The confining potential for a neutral atom in a standing optical wave
${\bf E}({\bf r},t)={\bf E}_0({\bf r}) {\rm Re} [\exp{(-i\omega_{\rm las}t)}]$ is 
$V_{\rm conf}({\bf r})=-(\varepsilon_0/4)\alpha'{\bf E}_0^2({\bf r})$,
 where $\alpha'=-e^2/(2m_e\omega_{\rm las}\varepsilon_0\Delta)$ is 
 the real part of the polarizability \cite{Grynberg}. Let
  us consider a red-detuned laser field corresponding 
  to $\Delta<0$ and $\alpha'>0$. In this configuration, 
  the atoms are  trapped around the maximum of the 
  electric field. For a mixture of two species, each species
  experiences its own 
detuning $\Delta_{\rm K}$ ($\Delta_{\rm Rb}$) given by the two transition 
wavelengths   $\lambda_{\rm K}=767$ nm and $\lambda_{\rm Rb}=780$ nm.
Within a parabolic approximation for the potential  around its
minimum, the ratio
$\omega_{\rm K}/\omega_{\rm Rb}$ of trap frequencies for K and 
Rb atoms becomes
\begin{equation}
\frac{\omega_{\rm K}}{\omega_{\rm Rb}}=\left(\frac{\Delta_{\rm Rb}}
{\Delta_{\rm K}}\frac{m_{\rm Rb}}{m_{\rm K}}\right)^{1/2}\, .
\end{equation}
Let us estimate this ratio for typical parameters.
In order to suppress spontaneous emission, we
  assume an average detuning of $\Delta=(\Delta_{\rm K}+\Delta_{\rm Rb})/2=-0.1
  \omega_{\rm las}$, yielding  $\omega_{\rm las}=5hc(\lambda_{\rm K}^{-1}+
\lambda_{\rm Rb}^{-1})/11 $ and $\Delta_{\rm Rb}/\Delta_{\rm k}=
(5\lambda^{-1}_{\rm K}-6\lambda^{-1}_{\rm Rb})/(5\lambda^{-1}_{\rm Rb}-6\lambda^{-1}_{\rm K})=0.84$. 
Taking also into account the mass ratio 
$m_{\rm Rb}/m_{\rm K}=87/40$, we have $\omega_{\rm K}/\omega_{\rm Rb}=1.35$, 
indicating a substantial coupling of COM and relative degrees of freedom. 

\begin{figure}[h]
\begin{center}
\epsfig{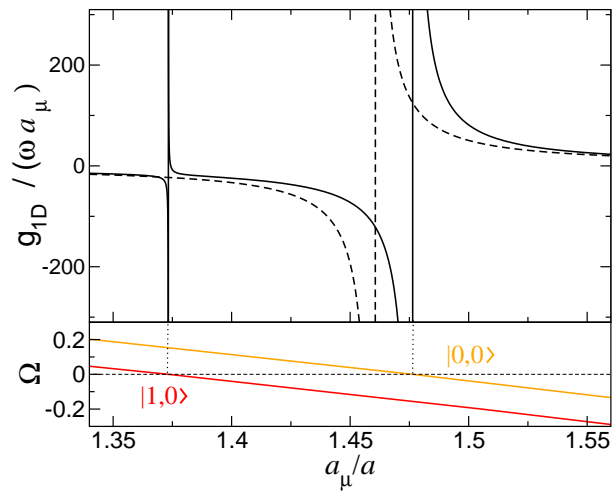}
\caption{Upper viewgraph: 
CIR in the effective interspecies 1D interaction constant 
$g_{\rm 1D}$ as a function of $a_{\mu}/a$. We consider a two-component 
atom gas of $^{40}$K and $^{87}$Rb, with average detuning 
$\Delta=-0.1 \omega_{\rm las}$ (solid line). For
comparison, we also show the result for the case when 
the two species experience the same trap frequency (dashed line). Lower viewgraph: 
 Dimensionless binding energy $\Omega$ for the two states $|0,0\rangle$
 and $|1,0\rangle$. \label{fig.2}}
\end{center}
\end{figure}

Using Eq.\  (\ref{G^2b}), we can project the Green's function  
$G_t({\bf R}_\perp,0;{\bf R}^\prime_\perp,0)$ on the appropriate 
basis defined in Eq.\ (\ref{deff}) and then compute numerically 
$\tilde{\zeta}_E$ 
by performing the imaginary-time integration, see Appendix B. 
Then $\tilde{\zeta}_{E_0}$ can be diagonalized, and the
effective interspecies 1D interaction constant $g_{\rm 1D}$ follows according to Eq.\
(\ref{u0}). The results are shown in the upper viewgraph of 
Fig.\  \ref{fig.2} in terms of
the characteristic length $a_{\mu}=\sqrt{2/(\mu 
(\omega_{\rm K}+\omega_{\rm Rb}))}$. We find two resonances,
 indicating that the discussion of Sec.\  \ref{remarks} applies to this particular 
 case. In order to illustrate the
interpretation of the CIR in terms of  Feshbach-type
resonances with bound states of the closed channels, we also plot 
in the lower viewgraph of Fig.\  \ref{fig.2} the
dimensionless binding energy $\Omega=2(E-\omega_{\rm K}+\omega_{\rm Rb})/
(\omega_{\rm K}+\omega_{\rm Rb})$ of the corresponding bound 
state.  As expected, the
resonances occur at those values of $a_{\mu}/a$ for which the energy
of the bound state of the closed channels coincides with the continuum
threshold of the open channel.

\section{Non-parabolic confining potentials\label{sec.nonlinear}}
Describing the potential created by an optical or a magnetic guide as parabolic
 is to some extent a simplification which has to be verified. In fact, even 
 though  the lower-energy transverse states can rather well 
 be approximated by the eigenfunctions of a
 2D harmonic oscillator, in every real trap the
 confinement is to some degree non-parabolic. For resonant
 scattering, we expect to have a virtual occupation 
 of many non-parabolic transverse 
states.  As a consequence, the location of the CIR will be slightly
moved, and new resonances could be created. This can already be seen
from an analysis similar to the one in Sec.\  \ref{remarks} for
small non-parabolic corrections. In order to tackle the problem
quantitatively, a full numerical treatment is required since no
analytical expression for the Green's function is in general 
available, in contrast to  Sec.\ \ref{sec.parabolic}.  

As an example, we consider the small non-parabolicity due to the
presence of a longitudinal magnetic bias field $B_z$ in a magnetic waveguide
containing a single-species gas. 
This is necessary to avoid Majorana spin flips \cite{Sukumar} 
and the subsequent escape of atoms out of the trap. 
 A magnetic trapping potential is formed according to $V_{\rm
 conf}({\bf x})=\mu_m 
|{\bf B}({\bf x})|$, where ${\bf B}({\bf x})$ is the applied magnetic
field and $\mu_m=m_F g_F \mu_B$, with $m_F$ being the
magnetic quantum number of the atom in the hyperfine state 
$|F, m_F\rangle$, $g_F$ the Land\'{e} factor and $\mu_B$ the Bohr
magneton. 
Assuming that apart from the longitudinal bias field, 
the remaining magnetic fields create a parabolic and isotropic
confinement in the transverse direction, the total
confinement is given by 
\begin{equation}\label{poten}
\chi V_{\rm conf}({\bf x}) = \sqrt{1+2 \chi (x^2+y^2)} \, , 
\end{equation}
 where we have scaled energy in units of the parabolic trapping
 frequency $\omega$ and  length in units of  
 $a_\mu= (\mu\omega)^{-1/2}$. The parameter $\chi=\omega/(\mu_m B_z)$
is related to the Majorana spin flip rate $\Gamma_{\rm loss}$
 \cite{Sukumar}. 
 The 1D effective interaction strength $g_{\rm 1D}$ can be calculated
 following our general approach. We compute $\tilde{\zeta}_{E_0}$
 numerically as outlined in Appendix B. 
 The  results are shown in Fig.\ \ref{fig.3} for $\chi=0.067$, which
 corresponds to $\Gamma_{\rm loss}=10^{-6}\omega$. We find two 
 resonances reflecting the cylindrical symmetry of the potential 
  (\ref{poten}) and the weakness of non-parabolic 
 corrections. The degeneracy of the parabolic case (shown in Fig.\ 
 \ref{fig.3}) is lifted and the original CIR is split into
 two nearby resonances. As expected, the effect of the
 non-parabolic transverse states shows up only 
 in the deep resonant region, making the parabolic solution a very
 good approximation away from the resonant region.
 In turn, this requires a good
 experimental resolution in order to observe the two CIR.
  
\section{Conclusions\label{sec.conclusio}}
To conclude, we have presented the general solution for
two-body s-wave scattering in a two-component ultracold atom gas
longitudinally confined to  one dimension by an arbitrary trapping
potential. The underlying key property is that the center-of-mass and the
relative degrees of freedom of the two-particle problem do not decouple, as
it is the case for a one-component gas and a 
 pure parabolic confinement. Thus, no reduction to an
effective single-particle problem is possible and the full coupled system
has to be solved. In the framework of the pseudopotential approach, 
we derive the energy of the bound state 
when all transverse channels are closed. Simple analytical
 results were obtained in 
the limiting cases of the dimer as well as the BCS limit. Moreover, 
 scattering solutions have been obtained when just one transverse
 channel is open.  The effective 1D interaction constant $g_{\rm 1D}$ can
 be calculated after diagonalizing a reduced Green's function. This can be
 achieved analytically for the special case of parabolic
 confinement, where the well-known confinement-induced resonance is recovered.
 For a two-component gas, as well as for a non-parabolic confinement, more
 than one CIR occur, which reflect the symmetry properties of the
 confining potential. These findings were illustrated by applying our
 formalism to experimentally relevant questions. We are confident
 that once the CIR has been verified experimentally, also
 the effects of a non-parabolic trapping potential will be discerned.

\section*{Acknowledgments}
We thank A.\ Gogolin  and A. G\"orlitz
for discussions. This work has
been supported by the DFG-SFB/TR 12.

\section*{Appendix A: Short-time 
 Green's function \label{app.1}}
In this Appendix, we illustrate how to expand the Green's function
\begin{equation}
 G_t({\bf X};{\bf X}')=
\langle{\bf X}|\exp{[-(K(\boldsymbol{\Pi})+U({\bf X}))t]}|{\bf X}'\rangle
\end{equation}
with respect to $t$ yielding the expression in Eq.\ (\ref{smallt}) for 
$G_t({\bf R}_\perp,0;{\bf R}_\perp',0)$. 
 In order to simplify the notation, we have introduced the 
five-dimensional vectors ${\bf X}=\{{\bf R}_\perp,{\bf r}\}$ and 
 $\boldsymbol{\Pi}=\{{\bf P}_\perp,{\bf p}\}$ and 
the functions  $K(\boldsymbol{\Pi})={\bf P}_\perp^2/2M+
{\bf p}^2/2\mu$ and
 $U({\bf X})=V_1({\bf R}_\perp+\mu {\bf r}_\perp/m_1)+
 V_2({\bf R}_\perp- \mu {\bf r}_\perp/m_2)$ for the 
 kinetic and the potential energy, respectively.
 First, we expand the Green's function
around the free solution given by 
\begin{equation}\fl\langle{\bf X}|\exp{[-K(\boldsymbol{\Pi})t]}|
{\bf X}'\rangle=\frac{M}{2\pi t}
\exp{\left[-\frac{({\bf R}_\perp-{\bf R}_\perp')^2M}{2t}
\right]}\left(\frac{\mu}{2\pi t}\right)^{3/2}
\exp{\left[-\frac{({\bf r}-{\bf r}')^2\mu}{2t}\right]}\,.
\label{frees}\end{equation}
In order to justify such an expansion, note that 
for $t\to 0^+$
\begin{equation}
\fl\langle{\bf X}|K(\boldsymbol{\Pi})
\exp{[-K(\boldsymbol{\Pi})t]}|{\bf X}'\rangle
= -\frac{d}{dt}
\langle{\bf X}|\exp{[-K(\boldsymbol{\Pi})t]}|
{\bf X}'\rangle \propto  
\delta({\bf X}-{\bf X}')
\frac{1}{t} \, ,
\label{kinen}
\end{equation}
whereas
\begin{eqnarray}
\fl\langle{\bf X}|U({\bf X})
\exp{[-K(\boldsymbol{\Pi})t]}|{\bf X}'\rangle 
& = & U({\bf X})\langle{\bf X}|
\exp{[-K(\boldsymbol{\Pi})t]}|{\bf
X}'\rangle 
\propto U({\bf X})\delta({\bf X}-{\bf X}') \, .
\label{potener}
\end{eqnarray}
Since the kinetic energy in Eq.\ (\ref{kinen}) 
diverges whereas the potential energy in Eq.\  
(\ref{potener}) remains finite, the latter can be regarded as a small
 perturbation. 
This expansion yields
\begin{equation}\label{firstexp}
G_t({\bf X};{\bf X}')\simeq(1-tU({\bf X}))
\langle{\bf X}|\exp{[-K(\boldsymbol{\Pi})t]}|{\bf X}'\rangle \, .
\end{equation}
Let us now set ${\bf X}_0=\{{\bf R}_\perp',0\}$ in Eq.\ (\ref{frees}) 
and expand with respect to $t$:
\begin{eqnarray*}
\lefteqn{\langle{\bf R}_\perp,0|\exp{[-K(\boldsymbol{\Pi})t]}|
{\bf R}_\perp',0\rangle } \\
&=& 
\left(\frac{\mu}{2\pi t}\right)^{3/2}
\frac{M}{2\pi t}\exp{\left[-\frac{({\bf R}_\perp-{\bf
R}_\perp')^2M}{2t}\right]} 
\\&=&
\left(\frac{\mu}{2\pi t}\right)^{3/2}\int\frac{d^2{\bf P}_\perp}{(2\pi)^2}
\exp{[i{\bf P}_\perp\cdot({\bf R}_\perp-{\bf R}_\perp')-\frac{{\bf P}_\perp^2}{2M}t]}
\\
&\simeq&\left(\frac{\mu}{2\pi t}\right)^{3/2}\int
\frac{d^2{\bf P}_\perp}{(2\pi)^2}
\left(1-\frac{{\bf P}_\perp^2}{2M}t\right)\exp{[i{\bf P}_\perp
\cdot({\bf R}_\perp-{\bf R}_\perp')]}\nonumber\\
&=&\left(\frac{\mu}{2\pi t}\right)^{3/2}\left(
\delta\left({\bf R}_\perp-{\bf R}_\perp'\right)-t\frac{{\bf P}_\perp^2}{2M}\right)\,.
\end{eqnarray*}
In the last line, the operator ${\bf P}^2_\perp$ stands for 
$(2\pi)^{-2}\int d^2{\bf P}_\perp \langle {\bf R}_\perp
| {\bf P}_\perp \rangle {\bf P}^2_\perp\langle {\bf P}_\perp | {\bf
R}_\perp' \rangle$.
Inserting  ${\bf X}_0$ into Eq.\ (\ref{firstexp}), we finally obtain 
Eq.\ (\ref{smallt}).
\section*{Appendix B: Evaluation of the operators $\zeta_E$ and
$\tilde{\zeta}_E$ \label{app.2}}
In this Appendix, we outline the evaluation of the kernels $\zeta_E$ and
$\tilde{\zeta}_E$ given in Eqs.\  (\ref{kernel2}) and (\ref{kernel3}),
respectively. 
\subsection*{B1: Parabolic confinement, $\omega_1 \ne \omega_2$}
First, let us consider the special case of parabolic confinement, but
the two species may experience different trap frequencies. For this
 confinement,  the Green's function 
$G_t({\bf R}_\perp,0;{\bf R}^\prime_\perp,0)$ is given in Eq.\  
(\ref{G^2b}).  
The first step is to project this operator 
onto the appropriate orthonormal basis $\{|j\rangle\}$ defined in
 Eq.\ (\ref{deff}).
 Note that this definition allows an arbitrary choice of the  basis,
 apart from properly fixing the vector $|0\rangle$. 
 One possibility is 
 introduced in Eq.\ (\ref{lagpol}). This is a natural option  because
 it reflects the cylindrical symmetry of the problem. However, this choice
 would not permit further analytical progress. 
 For this reason, we employ an  alternative basis defined by 
\begin{equation}
\langle{\bf R}_\perp|j\rangle=\langle{\bf R}_\perp|n_x,n_y\rangle=
\frac{1}{a_{M}}\psi_{n_x}\left(\frac{x}{a_{M}}\right)
\psi_{n_y}\left(\frac{y}{a_{M}}\right) \, ,
\end{equation}
where $\psi_n(x)$ is the eigenfunction for the 1D oscillator in 
dimensionless units, 
%
$\psi_n(x)=(\sqrt{\pi}2^{n_x}n!)^{-1/2}
\exp(-x^2/2)
H_{n}\left(x\right)
$, 
%
with  $H_{n}\left(x\right)$ being Hermite polynomials. Note that
the $x$ and
$y$ directions  factorize in the Green's function  (\ref{G^2b}),
allowing to perform the $x$ and
$y$ integrals separately. For convenience, we  introduce 
dimensionless coordinates $x \rightarrow x/a_{M}$  and find
\begin{eqnarray}\label{greeapp}
[G(t)]_{{\bf n},{\bf m}}&=&\langle n_x,n_y|G_t({\bf R}_\perp,0;
{\bf R}^\prime_\perp,0)|m_x,m_y\rangle\nonumber\\&=&
\sqrt{\frac{\mu}{2\pi t}}\frac{\beta(1-\beta)}{\pi^2 a_{M}^2}
\frac{e^{-\omega_1 t}}{1-e^{-2\omega_1 t}}
\frac{e^{-\omega_2 t}}{1-e^{-2\omega_2 t}}[F(t)]_{n_x,m_x}[F(t)]_{n_y,m_y}
\end{eqnarray} 
with
 \begin{equation}
[F(t)]_{n,m}=\int\,dxdx'\bar{\psi}_{n}\left(x\right)\exp{\left[-
\frac{x^2+x'^2}{2}f(t)+
xx'g(t)
\right]}\psi_{m}\left(x'\right).
\end{equation}
The functions $f(t)$ and $g(t)$ are defined in Eq.\  (\ref{fgfunc}). 
We perform the first integration by using the identity \cite{Grad}
\begin{equation}
\int dz \, e^{-(z-z')^2}H_n(\alpha z)=\pi^{1/2}(1-\alpha^2)^{n/2}
H_n\left(\frac{\alpha z'}{(1-\alpha^2)^{1/2}}\right)\,,
\end{equation}
with $\alpha=\alpha(t)=[(1+f(t))/2]^{-1/2}$, $z=x/\alpha(t)$ and 
$z'=g(t)\alpha(t)x'/2$, yielding 
\begin{eqnarray}\label{foft}
[F(t)]_{n,m}= (2^{n+m}m!n!)^{-1/2}\alpha(t)(1-\alpha(t)^2)^{n/2}
\nonumber\\\fl\times
\int dx'\exp{\left[-x'^2\left(\alpha^{-2}(t)-\frac{g(t)\alpha^2(t)}{4}
\right)\right]}H_n\left(
\frac{g(t)\alpha^2(t)}{2(1-\alpha^2(t))^{1/2}}x'\right)H_m(x')\, .
\end{eqnarray}
 By substituting Eq.\ (\ref{greeapp}) together with Eq.\  
 (\ref{foft}) into  $\zeta_E$ defined
 in Eq.\ (\ref{kernel2}), and by introducing the dimensionless time 
 $t'=\sqrt{t(\omega_1+\omega_2)}$, we get 
\begin{eqnarray}\label{num}
[\zeta_E]_{{\bf n},{\bf m}}& = & \frac{1}{4\pi a_\mu}
\int_0^\infty dt'\,\left\{A\,h_E(t')
\left[ F\left(\frac{t'^2}{\omega_1+\omega_2}
\right)\right]_{m_x,n_x} \right.
\nonumber \\ 
& & \left. \times \left[F
\left(\frac{t'^2}{\omega_1+\omega_2}\right)\right]_{m_y,n_y}
-\frac{2}{\pi^{1/2}t'^{2}}\delta_{{\bf n},{\bf m}}\right\} 
\end{eqnarray}
with the dimensionless parameter
$A=2\pi^{-3/2}\beta(1-\beta)a_\mu^2/a_{M}^2$
and 
\begin{eqnarray}
h_E(t')&=&
\exp{\left[-\frac{(\omega_1+\omega_2-E)t'^2}{\omega_1+\omega_2}\right]}
\left(1-\exp{\left[-\frac{2\omega_1t'^2}{\omega_1+\omega_2}\right]}\right)^{-1}
\nonumber \\ 
& & \times \left(1-\exp{\left[-\frac{2\omega_2t'^2}{\omega_1+\omega_2}\right]}\right)^{-1}.
\nonumber \\
\end{eqnarray}
It is now possible to evaluate the matrix elements of 
 $[\zeta_E]_{{\bf n},{\bf m}}$
 by numerically computing the double integrals in Eq.\ (\ref{num}).
 Note that the integrand does not suffer
 from any singularity due to the rescaling of the integration
 variable. Moreover, the convergence of the $x'$ integral (\ref{foft}) is
 exponentially fast. The first term in the integrand of the 
 $t'$ integral decays 
 exponentially at large times. Hence for large times, 
  only the second term yields a contribution, where the integration can
  be performed analytically in this region. 
For the case of interspecies scattering of Rb and K in an optical
trap, all the parameters entering in $A, h_E(t)$ and $[F(t)]_{n,m}$ 
can be expressed in terms of the ratios 
 $m_{\rm Rb}/ m_{\rm K}$ and $\Delta_{\rm Rb}/ \Delta_{\rm
K}$. The generalization to determine $\tilde{\zeta}_E$ is
straightforward and not detailed further.

\subsection*{B2: Non-parabolic confinement}
A numerical evaluation of the operator $\zeta_E$ and  $\tilde{\zeta}_E$
is less straightforward  when the Green's function 
$G_t({\bf R}_\perp,0;{\bf R}^\prime_\perp,0)$ cannot be computed analytically. 
In this case, $G_t({\bf R}_\perp,0;{\bf R}^\prime_\perp,0)$ should be computed 
by numerical diagonalization of  the
 $H_{\perp,i}$ and inserting their eigenvalues and eigenfunctions
 into Eq.\ (\ref{G^2}). 
For large $t$, this is feasible because only a small number of
eigenfunctions
 contribute to the  sum. However, for $t\to 0$, the number of 
eigenvectors required to cancel the divergence in  Eq.\  (\ref{kernel2})
quickly proliferates. This practical limitation can fortunately be
circumvented by the following trick.
Let us formally rewrite Eq.\ (\ref{kernel2}) as
\begin{eqnarray}
\label{appzeta}
\zeta_{E}({\bf R}_\perp,{\bf R}_\perp^\prime) 
&=&\int_0^{\infty}\frac{dt}{2\mu}\,\Big\{e^{Et}
[G_t({\bf R}_\perp,0;{\bf R}_\perp^\prime,0)-G^0_t({\bf R}_\perp,0;{\bf R}_\perp^\prime,0)]
\nonumber\\
&&+e^{Et}G^0_t({\bf R}_\perp,0;{\bf R}_\perp^\prime,0)
-\left(\frac{\mu}{2\pi t}\right)^{3/2}
\delta({\bf R}_\perp-{\bf R}_\perp^\prime)\Big\}\nonumber\\
&=&\int_0^{\infty}\frac{dt}{2\mu}\,e^{Et}
[G_t({\bf R}_\perp,0;{\bf R}_\perp^\prime,0)-G^0_t({\bf R}_\perp,0;{\bf R}_\perp^\prime,0)]
\nonumber\\
&&+\zeta^0_{E}({\bf R}_\perp,{\bf R}_\perp^\prime)\, , 
\end{eqnarray}
where $G^0_t({\bf R}_\perp,0;{\bf R}_\perp^\prime,0)$  and 
$\zeta^0_{E}({\bf R}_\perp,{\bf R}_\perp^\prime)$  are the
 Green's function and the integral kernel,  
respectively,  for an arbitrary reference confining potential
$V_0({\bf x}_\perp)$. If 
$G^0_t({\bf R}_\perp,0;{\bf R}_\perp^\prime,0)$  is known 
 analytically, we can deal with 
$\zeta^0_{E}({\bf R}_\perp,{\bf R}_\perp^\prime)$ as 
in the previous section. For confining potentials close to the
parabolic case, we choose a parabolic  $V_0({\bf x}_\perp)$.

Regarding Eq.\  (\ref{appzeta}), we proceed as follows.
We restrict the infinite-dimensional Hilbert space to the ${\cal N}$
lowest eigenstates of the potential $V_0({\bf x}_\perp)$, and
diagonalize the original Hamiltonian in this ${\cal N}$-dimensional 
Hilbert space. With the eigenfunctions at hand, the Green's function
can be computed using Eq.\  (\ref{G^2}). Then, the sum in  Eq.\ 
(\ref{G^2}) is exchanged with the $t$-integration and the latter is
performed. Next, we project the Green's function onto a known 
single-particle basis $\{|m\rangle\}$.
 To achieve numerical 
convergence,  we increase the Hilbert space dimension ${\cal N}$ until
the result does not change anymore.
We emphasize that the overall result converges to the exact result although obviously 
not all the single-particle  states used in computing the Green's
function are reliable on very long distances (comparable to the
numerical system size) because higher-lying energy states are
increasingly inaccurate. Nevertheless, the central part (in position space) of the
eigenfunctions -- which corresponds to the kinetic energy and  does
not feel the confinement -- is accurate enough to cancel the divergence
stemming from the kinetic part. In order to compute the scattering
solution, we compute $\tilde{\zeta}_{E_0}$ with an analogous procedure,
diagonalize $\tilde{\zeta}_{E_0}$ numerically, and insert the result into
Eq.\  (\ref{u0}). 
For the non-parabolic confinement in Sec.\ 
\ref{sec.nonlinear}, a
 parabolic   $V_0({\bf x}_\perp)$ is appropriate.
 In this case, we use for $\{|m\rangle\}$
 the orthonormal basis  defined in Eq.\ (\ref{lagpol}). 
 Then $\zeta^0_{E}$ is diagonal and 
given by Eq.\ (\ref{Zetapar}). 

\vspace*{15mm}

\begin{figure}[h]
\begin{center}
\epsfig{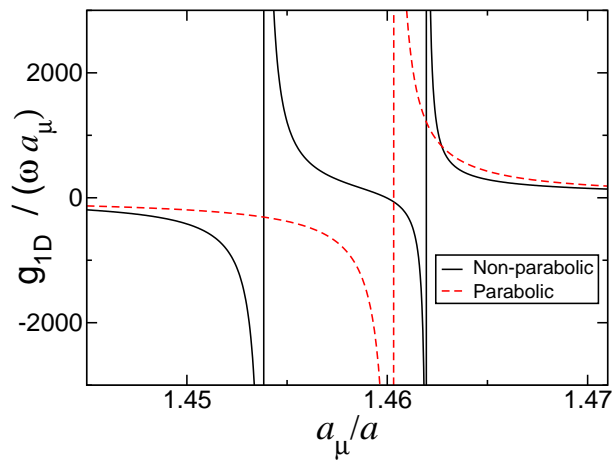}
\caption{1D effective interaction strength $g_{\rm 1D}$ for 
the non-parabolic potential of Eq.\ (\ref{poten}) (black solid line) for $\chi=0.067$ 
 corresponding to $\Gamma_{\rm loss}=10^{-6}\omega$. For comparison,
 the parabolic case is also shown (red dashed line). 
 \label{fig.3}}
\end{center}
\end{figure}
\end{document}